# nvidia-pcm: A D-Bus-Driven Platform Configuration Manager for OpenBMC Environments

Harinder Singh, Senior Member, IEEE

*NVIDIA Corporation, Santa Clara, CA, USA*

Email: harinders@nvidia.com

ORCID: 0009-0005-9235-124X

*Abstract—*

GPU-accelerated server platforms that share most of their hardware architecture often require separate firmware images due to minor hardware differences—different component identifiers, thermal profiles, or interconnect topologies. I built nvidia-pcm to eliminate that overhead. nvidia-pcm is a platform configuration manager for NVBMC, NVIDIA's OpenBMC-based firmware distribution, that enables a single firmware image to serve multiple platform variants. At boot, nvidia-pcm queries hardware identity data over D-Bus and exports the correct platform-specific configuration as environment variables. Downstream services read those variables without knowing or caring which hardware variant they are running on. The result is that platform differences are captured entirely in declarative JSON files, not in separate build artifacts. This paper describes the architecture, implementation, and deployment impact of nvidia-pcm, and shares lessons learned from solving the platform-identity problem at a deliberately minimal level of abstraction—prioritizing adoption simplicity over comprehensive hardware modeling.

*Actionable Insights*

• A single shared firmware image can serve multiple hardware variants if platform identity is resolved at runtime and injected through a centralized service. This eliminates the per-variant build, test, and release overhead that dominates firmware maintenance costs as platform portfolios grow.

• Declarative JSON configuration files that separate platform-specific knowledge from service code let engineers add support for a new hardware variant by dropping one file into an existing image—no code changes, no recompilation, no new build pipeline.

• Environment variables written to a standard file (/etc/default/nvidia-pcm) and consumed via systemd's EnvironmentFile= directive required no code changes in downstream services—only a one-line addition to each service's unit file. Choosing this deliberately simple mechanism over D-Bus properties or structured config files proved critical to adoption across teams.

## I. INTRODUCTION

Enterprise servers include a Baseboard Management Controller (BMC)—an embedded processor running its own Linux-based firmware, independent of the host operating system. The BMC handles out-of-band management functions: power control, hardware health monitoring, remote console access, and firmware lifecycle management. Software engineers working on infrastructure-as-code or platform engineering will recognize these as the same operational concerns they manage at the fleet level, but implemented per-machine in dedicated firmware.

For GPU-accelerated AI servers, the BMC's job is especially complex. These platforms combine multiple GPUs, high-bandwidth interconnects like NVLink, specialized power delivery, and aggressive thermal management. The BMC firmware needs to know which specific hardware variant it is running on so that it can configure GPU management daemons with the right topology manifests, point error-handling services to the correct event definitions, and set interconnect monitoring to the right link profiles.

The problem nvidia-pcm addresses is not just code duplication—it is that platforms sharing most of the same hardware architecture required separate firmware images. Each image meant a separate build pipeline, a separate test cycle, and separate release management. For platforms that differed only in minor variants—a different FRU identifier, a slightly different GPU count, a different thermal profile—maintaining entirely separate images was unsustainable as the platform portfolio grew.

nvidia-pcm makes the firmware image platform-agnostic. Each platform variant is described in a JSON configuration file that specifies what D-Bus properties to check and what values to expect—for example, matching the PRODUCT_PRODUCT_NAME field exposed by the FRU reader service (xyz.openbmc_project.FruDevice). At boot, nvidia-pcm iterates through these configuration files, queries D-Bus (the IPC bus used by OpenBMC services) for the specified properties, and when a match is found, exports the corresponding environment variables. Platform-specific differences live entirely in these JSON files, not in separate builds. The services that depend on this configuration read their environment variables without any awareness of platform identity.

nvidia-pcm runs within NVBMC, NVIDIA's OpenBMC-based firmware distribution. This paper presents the work as a case study because the underlying problem—too many build artifacts for too-similar platforms—is not unique to

NVIDIA. Any team managing a growing portfolio of hardware variants on OpenBMC or similar embedded Linux firmware faces the same pressure.

This paper contributes:
- A production case study showing how runtime platform detection replaced per-variant firmware images in an NVIDIA BMC environment.
- The architecture and implementation of nvidia-pcm: declarative JSON configurations, D-Bus-based hardware queries, and environment variable export.
- Lessons learned from deploying centralized configuration management in production BMC firmware.

## II. BACKGROUND AND MOTIVATION

*A. OpenBMC and D-Bus*

OpenBMC is an open-source Linux distribution purpose-built for BMCs. Its key architectural decision is modularity: rather than a single monolithic firmware binary, OpenBMC distributes functionality across independent services that communicate through D-Bus [8], a system message bus widely used in Linux environments. Each service registers objects on D-Bus, exposing properties that other services can read. An ObjectMapper service keeps track of what is registered where, so services can discover hardware resources at runtime without hardcoded paths.

Each hardware component carries identity data in a Field Replaceable Unit (FRU) record, stored in a small EEPROM (Electrically Erasable Programmable Read-Only Memory) chip. Dedicated FRU reader services—such as xyz.openbmc_project.FruDevice—scan these EEPROMs over I2C and expose the parsed fields (product name, manufacturer, serial number) as D-Bus properties. nvidia-pcm does not access hardware directly; it queries these D-Bus properties, matching them against expected values defined in its JSON configuration files.

*B. The Firmware Image Problem*

The immediate pain point was not detection code—it was firmware images. GPU-accelerated platforms in the same product family often share 90% or more of their hardware design. The GPU baseboard is the same; the BMC SoC is the same; most of the firmware services are identical. But minor differences—a different baseboard FRU identifier, different GPU count, different NVLink topology—meant each variant needed its own firmware build. Each build carried its own test matrix, its own release cycle, and its own field-update logistics.

As the number of platform variants grew, this approach did not scale. The engineering cost was dominated not by the firmware code itself but by the build and release infrastructure surrounding it. The team needed a way to ship one image that could adapt to whichever hardware variant it booted on.

Prior to nvidia-pcm, NVBMC did not perform runtime platform detection—platform identity was determined at build time by producing separate firmware images. Services did not need detection logic because each image was already tailored to one specific platform. While the broader OpenBMC ecosystem offers runtime hardware discovery through frameworks like Entity Manager [4], NVBMC required a lighter-weight solution focused specifically on platform-variant identification and configuration export. nvidia-pcm introduced runtime detection as a replacement for the per-image approach, centralizing it in a single service so that individual services never need to be aware of platform identity.

*C. Requirements*

nvidia-pcm needed to do three things: identify the platform from hardware-embedded data (not from a build-time flag), export the correct platform-specific configuration so services consume it without awareness of platform identity, and make it possible to add new similar platforms without rebuilding the firmware image.

## III. DESIGN GOALS AND REQUIREMENTS

*A. D-Bus-Native Platform Detection*

I designed nvidia-pcm to leverage the existing D-Bus infrastructure for hardware discovery rather than implementing direct hardware access:
- Abstraction from hardware details: D-Bus services like FruDevice already handle I2C communication and EEPROM parsing. nvidia-pcm reads the resulting properties without coupling to specific hardware access mechanisms.
- Dynamic discovery: The ObjectMapper's GetSubTree method lets nvidia-pcm discover all objects implementing a particular interface, so it works even when FRU objects appear at different D-Bus paths across platform variants.
- Consistency: nvidia-pcm reads the same D-Bus properties available to every other service on the BMC, ensuring identification decisions are based on the same data system-wide.

*B. Declarative Configuration*

I express platform definitions declaratively in JSON configuration files rather than procedurally in code. Each configuration file specifies:
- A platform name serving as the configuration's identifier
- A set of checks (D-Bus property queries with expected values)
- A set of actions (environment variables to export when checks pass)

System integrators can define configurations for their platform variants as supplementary JSON files, extending support without modifying the nvidia-pcm codebase or rebuilding the firmware image.

*C. Flexible Matching Rules*

I implemented two matching rules: MatchAll (all checks must pass) and MatchOne (any check passing suffices). The rule can be specified at both the configuration level and within individual check groups.

*D. Environment Variable Export*

I chose environment variables as the output mechanism, writing platform configuration to /etc/default/nvidia-pcm. I considered exposing the results as D-Bus properties or writing structured configuration files, but environment variables won on simplicity: any process can read them, services reference the file through the EnvironmentFile= option in their systemd unit configuration, and the resulting file is plain text that engineers can inspect directly during debugging. The file persists across service restarts, so services that start later still get the right configuration.

*E. Graceful Fallback*

When no platform-specific configuration matches, nvidia-pcm applies a default configuration rather than leaving the environment file empty. This keeps the system functional—if degraded—on unrecognized hardware, which matters during early hardware bring-up when FRU data may be incomplete.

## IV. SYSTEM ARCHITECTURE

*A. Architectural Overview*

nvidia-pcm runs as a oneshot systemd service during BMC initialization. It identifies the platform, writes configuration to the environment file, and exits. Every BMC service that starts afterward reads that file.

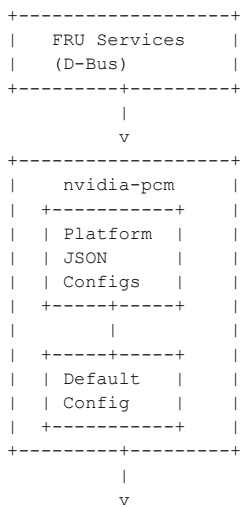

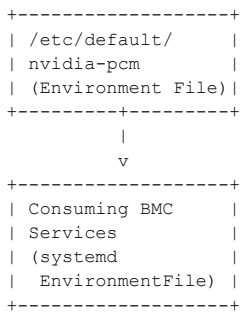

**Figure 1: nvidia-pcm Architecture**

*B. Platform Configuration Files*

Platform configurations are defined in JSON files stored in /usr/share/nvidia-pcm/platform-configuration-files/. Each file defines a single platform configuration:

```
{
    "Name": "Example Platform",
    "Checks": [
        {
            "rule": "MatchAll",
            "objects": [],
            "interface": "xyz.openbmc_project.FruDevice",
            "property": "PRODUCT_PRODUCT_NAME",
            "value": "Example Product Name"
        }
    ],
    "Actions": [
        {
            "variables": [
"CONFIG_MANIFEST=/usr/share/example/manifest.json",

"CONFIG_PROFILE=/usr/share/example/profile.json"
            ]
        }
    ]
}
```

Name is written to the environment file as NAME. Checks specifies D-Bus property queries: each check includes a rule, objects (empty triggers dynamic discovery), interface, property, and expected value. Actions lists environment variable assignments to write when checks pass.

*C. D-Bus Property Discovery*

When the objects array is empty, nvidia-pcm uses ObjectMapper's GetSubTree to discover all D-Bus objects implementing the specified interface. The discovery process filters results to objects owned by recognized FRU services (com.Nvidia.FruManager or xyz.openbmc_project.FruDevice), ensuring properties are read from authoritative sources.

*D. Check Execution*

nvidia-pcm processes configuration files iteratively: load each file, execute all checks, and if all pass, execute actions and exit. The first matching configuration wins. If no configuration matches, the default configuration is applied.

*E. Action Execution*

When a configuration matches, nvidia-pcm writes environment variables to /etc/default/nvidia-pcm with permissions 0664. The NAME variable is always written first, followed by platform-specific variables.

*F. Skip-Checks Optimization*

A --skip-checks mode accelerates subsequent boots: if the environment file already exists, nvidia-pcm reads the NAME variable, finds the matching configuration by name, and re-executes only the actions—skipping D-Bus queries entirely.

## V. IMPLEMENTATION

*A. Technology Stack*

I implemented nvidia-pcm in C++20, aligning with the OpenBMC ecosystem where most services use C++. The build system uses Meson. Dependencies include sdbusplus [1] for D-Bus bindings, phosphor-logging [2] for OpenBMC logging, and nlohmann/json [3] for JSON parsing.

*B. D-Bus Access Implementation*

I encapsulate D-Bus operations in the dbus namespace:
```
void getProperty(const std::string& service,
                 const std::string& objectPath,
                 const std::string& interface,
                 const std::string& property,
                 DBusValue& value);

DBusSubTree getSubTree(const std::string& interface);
```

*C. Configuration Loading*

The platform_config::Config class represents a loaded configuration:
```
class Config {
public:
    std::string name;
    std::string rule;
    std::vector<platform_checks::Checks_t> checks;
    std::vector<platform_actions::Actions_t> actions;

    bool loadFromFile(const std::string& file);
    bool performChecks();
    int performActions();
};
```

*D. Check Execution Algorithm*

```
Algorithm: Platform Check Execution
Input: Config object with checks array
Output: Boolean indicating match success

1. For each check in checks:
   a. If check.objects is empty:
      i.   Call getSubTree(check.interface)
      ii.  Filter results to FruManager/FruDevice services
      iii. Populate check.objects with discovered paths
   b. If check.objects is still empty:
      i.   Return false (no objects to check)
   c. For each object in check.objects:
      i.   Call getProperty(service, object, interface, property)
      ii.  Store result in check.dbusPropertyValues
   d. Apply check.rule:
      - MatchAll: All values must equal check.value
      - MatchOne: At least one value must equal check.value
   e. If check fails, return false
2. Return true (all checks passed)
```

*E. Action Execution Implementation*

```
int Actions_t::performActions(const std::string& name, bool& fileCreated) {
    std::ofstream envFile;
    if (fileCreated) {
        envFile.open(PCM_ENV_FILE, std::ofstream::out | std::ofstream::app);
    } else {
        envFile.open(PCM_ENV_FILE, std::ofstream::out | std::ofstream::trunc);
        fileCreated = true;
        fs::permissions(PCM_ENV_FILE, fs::perms::owner_write |
                         fs::perms::owner_read |
            fs::perms::group_read |
                         fs::perms::group_write |
            fs::perms::others_read);
    }
    envFile << "NAME=" << name << std::endl;
    for (const auto& variable : variables) {
        envFile << variable << std::endl;
    }
    envFile.close();
    return 0;
}
```

## VI. INTEGRATION AND DEPLOYMENT

*A. Service Integration via systemd*

BMC services consume nvidia-pcm output through systemd's EnvironmentFile directive:
```
[Unit]
Description=Example BMC Service
After=nvidia-pcm.service
Requires=nvidia-pcm.service

[Service]
```

```
EnvironmentFile=/etc/default/nvidia-pcm
ExecStart=/usr/bin/example-service --
manifest=${CONFIG_MANIFEST}
```

*B. Runtime Environment Access*

Services can also access exported variables directly:
```
const char* manifest =
std::getenv("CONFIG_MANIFEST");
if (manifest) { loadManifest(manifest); }
```
```
import os
manifest = os.environ.get('CONFIG_MANIFEST')
```

*C. Adding New Platform Support*

Adding a new platform variant that shares the same base architecture requires one JSON file placed in the configuration directory. No recompilation, no new firmware build, no new test pipeline. The existing image handles it at the next boot.

*D. Deployment Impact*

Due to the proprietary nature of the production deployment, I report qualitative and relative improvements rather than absolute figures.

Before nvidia-pcm, each new platform variant in the same product family required a dedicated firmware image. Even when the platforms shared nearly all of their hardware design and firmware code, the minor differences—a different FRU product name, different GPU topology manifest, different error-handling event table—meant a separate build artifact. Each image carried its own build pipeline, test matrix, and release cycle.

After nvidia-pcm, those variants share a single firmware image. Platform-specific behavior is captured entirely in JSON configuration files baked into the image. When nvidia-pcm runs at boot, it detects the hardware variant and exports the right configuration. Adding a new variant within the same product family means adding one JSON file to the existing image—not forking the build.

Complexity analysis. Without nvidia-pcm, supporting P platform variants across S services requires P separate firmware images, each carrying its own build pipeline, test matrix, and release cycle. Adding one new variant means creating one new image and its associated infrastructure. With nvidia-pcm, supporting P variants requires P configuration files within a single image. Adding one new variant requires one JSON file—no new build, no service modifications. The maintenance burden shifts from multiplicative (images × release cycles) to additive (one file per variant).

|  | Without nvidia-pcm | With nvidia-pcm |
|---|---|---|
| Supporting P variants | P firmware images | 1 image + P config files |
| Adding one new variant | New image + build pipeline + test cycle | One JSON file (~30 lines) |
| Adding one new service | Must be included in all P images | Added to one image; reads env vars |

Concrete example. As a representative case, two platform variants that differ only in their GPU topology manifest and error-handling event table share a single firmware image. The JSON configuration file that distinguishes them is approximately 30 lines. Without nvidia-pcm, these variants would have required separate firmware builds—identical in all other respects.

The degree of abstraction this created for downstream services turned out to be simpler than initially anticipated. The GPU management daemon, the error handler, the monitoring agents—none of them contain any platform detection logic. They read their manifest path or event table path from an environment variable, and they have no idea whether they are running on one platform variant or another. nvidia-pcm was built to solve a build-infrastructure problem, but the bigger payoff turned out to be this runtime abstraction: services became genuinely platform-agnostic, which made them simpler to develop, test, and maintain independently.

nvidia-pcm is available as open source at github.com/NVIDIA/nvidia-pcm under the Apache License 2.0.

**VII. DESIGN RATIONALE**

*A. Why D-Bus for Discovery*

Alternative approaches to platform identification include direct hardware access (reading FRU EEPROMs via I2C), file-based identification (configuration files provisioned during manufacturing), and DMI/SMBIOS queries. I chose D-Bus because OpenBMC already provides D-Bus services exposing FRU data; nvidia-pcm leverages this existing infrastructure rather than duplicating hardware access code. D-Bus provides a stable abstraction independent of I2C topology or EEPROM addressing, and ensures nvidia-pcm makes identification decisions based on the same data available system-wide.

*B. Why Environment Variables*

Alternative output mechanisms include D-Bus properties and writing platform-specific configuration files directly. I chose environment variables because they require no additional libraries, services reference the file through the EnvironmentFile= option in their systemd unit configuration, and the resulting file is plain text that engineers can inspect directly during debugging.

*C. Why First-Match Semantics*

A scoring or priority-based matching system would have been more flexible, but also harder to debug when it chose the wrong configuration. With first-match semantics, the behavior is deterministic and easy to reason about: if your platform is not matching, you check the configuration files in directory order until you find the one that should have matched. I also include a default fallback configuration so that an unrecognized platform gets some configuration rather than none—important during early hardware bring-up.

## VIII. CONSIDERATIONS AND LIMITATIONS

### A. FRU Data Quality

nvidia-pcm only works as well as the FRU data it reads. Not all hardware has fully populated FRU fields. Content can vary between manufacturing batches. Some FRU records contain encoding errors. In practice, I learned to target the one or two properties that the manufacturing process reliably populates and to avoid depending on fields that are sometimes left blank.

### B. Timing Dependencies

nvidia-pcm must start after the FRU service has finished populating D-Bus. I enforce this with systemd ordering dependencies (After=, Requires=). The skip-checks optimization helps on subsequent boots by avoiding D-Bus queries entirely when the platform identity is already known.

### C. Configuration File Ordering

Because nvidia-pcm uses first-match semantics and iterates over files in a directory, filesystem ordering matters. Configuration files follow a naming convention of plat_config_<platform>.json (e.g., plat_config_GB200.json) and are baked into the BMC firmware image, which is itself version-controlled.

### D. Security

nvidia-pcm runs inside the BMC's trusted execution environment. Configuration files are part of the read-only firmware image. FRU data integrity depends on the hardware EEPROM being trustworthy—a reasonable assumption for enterprise servers, but something to revisit if the threat model includes physical access attacks. The environment file is world-readable, which is fine for platform identity data that is not security-sensitive.

## IX. RELATED WORK

### A. OpenBMC Entity Manager

OpenBMC's Entity Manager [4] already provides runtime hardware discovery with JSON-driven configuration—its Probe mechanism detects hardware at boot via D-Bus, and its Exposes records drive downstream "reactor" daemons such as dbus-sensors. Entity Manager builds a complete topology model: one configuration file per supported device, D-Bus objects for every discovered component and their associations, and a schema that downstream services must implement as D-Bus-aware reactors. It is the established approach to runtime hardware configuration in the OpenBMC community.

nvidia-pcm solves a narrower problem. It does not model hardware topology, discover individual components, or create D-Bus objects. It answers one question—"which platform variant am I?"—and exports the answer as environment variables in a plain text file. Services consume this output through systemd's EnvironmentFile= option, requiring no D-Bus client code and no awareness of the detection mechanism. For the specific problem of mapping platform identity to service configuration files, nvidia-pcm provides a lighter path than adopting Entity Manager's full framework. The two tools can coexist: nvidia-pcm can consume D-Bus data that Entity Manager or simpler FRU services provide.

### B. Per-Service Detection

Most BMC firmware stacks I have encountered put the detection logic inside each service. This works when you have one or two platform variants and a handful of services. It breaks down when both numbers grow, because every new platform means touching every service [5]. nvidia-pcm follows a well-known architectural pattern: centralizing a cross-cutting concern into a dedicated service rather than distributing it across consumers [9].

### C. Configuration Management and Redfish

General-purpose configuration management tools (Ansible, Puppet, Salt) operate at the fleet level [6]. nvidia-pcm operates inside a single machine's BMC, at a much earlier stage in the boot process—before the network stack is fully up, let alone a fleet orchestrator. The DMTF Redfish specification [7] provides standardized APIs for platform management, but nvidia-pcm runs before Redfish services initialize and can feed platform identity into the Redfish stack once it starts [10].

## X. CONCLUSION

nvidia-pcm eliminates the overhead of maintaining separate firmware images for platform variants that share most of their hardware design. The tool detects the hardware variant at boot by querying FRU data over D-Bus, then exports the correct configuration as environment variables that downstream services consume without any awareness of platform identity.

Three things stood out from this work:

**The abstraction was the real payoff**. nvidia-pcm started as a solution to a build-infrastructure problem—too many firmware images for too-similar platforms. In practice, the runtime abstraction it provided was simpler and more effective than initially anticipated. Downstream services became genuinely unaware of which hardware variant they were running on. That made them simpler to write, simpler to test, and simpler to maintain. The pattern is analogous to feature flags or environment-based container configuration, but applied at a level of simplicity unusual in embedded firmware, where comprehensive hardware modeling frameworks like Entity Manager [4] are the more established approach.

**The simplest output mechanism was the right one**. Environment variables are about as unsophisticated as an IPC mechanism gets. That turned out to be exactly why they worked: no library dependencies, no protocol overhead, no learning curve. Every service team could integrate with nvidia-pcm by adding one line to a systemd unit file. If I had chosen a more complex mechanism, adoption would have been slower.

**The default fallback was essential for adoption**. Services need to start even on hardware that nvidia-pcm does not yet recognize—during early board bring-up, for example, when FRU data may be incomplete or not yet finalized. The default configuration ensures the system boots into a functional state rather than failing entirely, which made it practical to deploy nvidia-pcm before every platform variant had a dedicated configuration file.

nvidia-pcm is available as open-source software under the Apache License 2.0 at github.com/NVIDIA/nvidia-pcm.

## ACKNOWLEDGMENTS

I developed nvidia-pcm as part of NVIDIA's NVBMC firmware infrastructure. The project is available as open source at github.com/NVIDIA/nvidia-pcm under Apache License 2.0. Portions of this paper were revised with the assistance of AI-based language tools for improving clarity, structure, and tone. All technical content, architecture, implementation details, and deployment observations reflect the author's original work and direct experience.

## REFERENCES

[1] OpenBMC, "sdbusplus: C++ bindings for D-Bus." https://github.com/openbmc/sdbusplus

[2] OpenBMC, "phosphor-logging: OpenBMC logging infrastructure." https://github.com/openbmc/phosphor-logging

[3] N. Lohmann, "JSON for Modern C++." https://github.com/nlohmann/json

[4] OpenBMC, "Entity Manager." https://github.com/openbmc/entity-manager

[5] OpenBMC Project Documentation. https://github.com/openbmc/docs

[6] M. Burgess, "Cfengine: A Site Configuration Engine," *USENIX Computing Systems*, vol. 8, no. 3, 1995.

[7] DMTF, "Redfish Specification," DSP0266. https://www.dmtf.org/standards/redfish

[8] H. Pennington, A. Carlsson, A. Larsson, and S. Herzberg, "D-Bus Specification," freedesktop.org, 2023. https://dbus.freedesktop.org/doc/dbus-specification.html

[9] F. Buschmann, R. Meunier, H. Rohnert, P. Sommerlad, and M. Stal, "Pattern-Oriented Software Architecture, Volume 1: A System of Patterns," Wiley, 1996.

[10] DMTF, "Platform Level Data Model (PLDM) for Platform Monitoring and Control," DSP0248. https://www.dmtf.org/standards/pmci

## AUTHOR BIOGRAPHY

Harinder Singh is a Senior Software Engineer at NVIDIA Corporation, where he has spent 9 years architecting enterprise system management and platform reliability software for data center infrastructure. He is a Senior Member of IEEE. His work includes NVIDIA's system management platform (NVSM), reliability certification tooling (NVRASTool), and open-source BMC configuration management (nvidia-pcm). Contact him at harinders@nvidia.com.